\documentclass[preprint,prl]{revtex4}%
\usepackage{amsfonts}
\usepackage{amsmath}
\usepackage{amssymb}
\usepackage{graphicx}%
\providecommand{\U}[1]{\protect\rule{.1in}{.1in}}

\ifx\pdfoutput\relax\let\pdfoutput=\undefined\fi
\newcount\msipdfoutput
\ifx\pdfoutput\undefined\else
\ifcase\pdfoutput\else
\ifx\paperwidth\undefined\else
\ifdim\paperheight=0pt\relax\else\pdfpageheight\paperheight\fi
\ifdim\paperwidth=0pt\relax\else\pdfpagewidth\paperwidth\fi
\fi\fi\fi
\begin{document}
\title[Superstatistical Random Matrix Theory ]{Analysis of symmetry breaking in quartz blocks using superstatistical random
matrix theory }
\author{A.Y. Abul-Magd, S. A. Mazen and M. Abdel-Mageed}
\affiliation{Faculty of Science, Zagazig University, Zagazig, Egypt.}
\keywords{Random Matrix Theory, Superstatistics, acoustic resonances,Chaos}
\pacs{02.50.-r, 05.40.-a, 05.45.Mt, 05.45.Tp, 03.65.-w}

\begin{abstract}
We study the symmetry breaking of acoustic resonances measured by Ellegaard et
al. \cite{Ellegaard q} in quartz blocks. The observed resonance spectra show a
gradual transition from a superposition of two uncoupled components, one for
each symmetry realization, to a single component well represented by a
Gaussian orthogonal ensemble (GOE) of random matrices. We discuss the
applicability of superstatistical random-matrix theory to the final stages of
the symmetry breaking transition. A comparison is made between different
formula of the superstatistics and a pervious work \cite{Abd El-Hady}, which
describes the same data by introducing a third GOE component. Our results
suggest that the inverse-chi-square superstatistics could be used for studying
the whole symmetry breaking process.

\end{abstract}
\volumeyear{year}
\volumenumber{number}
\issuenumber{number}
\eid{identifier}
\date[Date text]{date}
\received[Received text]{date}

\revised[Revised text]{date}

\accepted[Accepted text]{date}

\published[Published text]{date}

\maketitle

\section{Introduction}

Few years ago Ellegaard et al. studied the influence of the flip symmetry
breaking in quartz blocks on their spectral statistics \cite{Ellegaard q}.
Crystalline quartz exhibits a $D_{3}$ point-group symmetry about the crystal's
$Z$ (optical) axis and three two-fold rotation symmetries about the three $X$
(piezoelectric) axes; the latter three axes lie in a plane orthogonal to the
$Z$ axis and sustain angles of 120$^{\circ}$ with respect to each other.
Ellegaard et al.\cite{Ellegaard q} used rectangular blocks of dimensions $14$
$\times\ 25$ $\times$ $40$ mm$^{3}$, cut in such a way that all symmetries are
fully broken except a two-fold `flip' symmetry about one of the three $X$
axes. In order to break the flip symmetry, Ellegaard et al removed an octant
of \ a sphere with a successfully increasing radius from one corner of the
rectangular block. The breaking of the flip symmetry causes the eigenmodes
belonging to the two different representations to interact and mix. As the
radius is made larger, the nearest-neighbor-spacing (NNS) distribution makes a
fast transition from a distribution of two non-interacting spectra towards the
distribution of one chaotic system without symmetries. Full chaos was attained
at an octant radius of $r=10$ mm, as the cut block had the shape of a three
-dimensional Sinai Billiard.

A problem like this is found, e.g., in nuclear physics where isospin symmetry,
characteristic of the strong interactions, is only approximate due to Coulomb
effects \cite{Mitchell}. Isospin mixing was analyzed by Guhr and
Weidenm\"{u}ller using a random matrix approach \cite{GuhrW}. They used a
random matrix model to describe experimental data and to estimate the average
symmetry-breaking matrix element, i.e., the average Coulomb matrix element.
The random matrix ensemble consists of matrices $H=H_{0}+V$, where $H_{0}$ is
block-diagonal, here two blocks for the two-fold symmetry, each a member of
the GOE; $V$ couples the blocks, breaking the symmetry. Level statistics were
studied within this model analytically by Leitner \cite{Leitner}. He assumed
that the probability density of spacing belonging to the same block are
unaffected by the coupling matrix $V$. He took into account the interaction
between pairs of levels belonging to different blocks, yielding linear
dependence of the spacing distribution near the origin. Taking into account
the interaction between more than two levels might lead to fractional-level
repulsion as recently shown by B\"{a}cker et al. \cite{backer}. While
Leitner's formalism is a perturbation-theory approach, which is valid only for
small $V$, he has used it as a model for the whole symmetry breaking process
in quartz-block data \cite{leitner}. El-Hady et al. \cite{Abd El-Hady}
described the transition by introducing a third GOE sequence which increased
in the expense of the initial two independent sequences. To fit the data in
Ref. \cite{Ellegaard q}, the authors assumed\bigskip\ that the initial
sequence are pseudointegrable (for definition, see the discussion following
Eq. (\ref{SP}) below). This independent-sequence model of symmetry breaking
was further elaborated in \cite{Magd}, where a relation between the fractional
level density of each sequence is related to the mean squared
symmetry-breaking matrix element. De Carvalho et al. \cite{Hussien} take into
account the interaction of the three level sequences using the perturbation
method suggested by Leitner \cite{Leitner}.

In the present paper we perform an analysis of data in Ref. \cite{Ellegaard q}
using a superstatistical generalization of Random Matrix Theory (RMT)
\cite{RMT,Brody,Bohigas,Gur,Dyson}. We are motivated by the success of the
superstatistical approach to describe the final stages of the transition from
integrability to chaos. A brief account of this approach is given in Section
2. Section 3 reports the results of the superstatistical analysis of symmetry
breaking. We also compare the results our work with that of Ref. \cite{Abd
El-Hady}. A discussion of the results of this contribution is given in Section 4.

\section{Superstatistical RMT}

The concept of superstatistics was introduced by Beck and Cohen \cite{C C} to
describe deviation of thermodynamic system from equilibrium. This concept has
been successfully applied to a wide variety of physical problems, including
turbulence \cite{C R}, plasma physics \cite{Sattin}, cosmic-ray statistics
\cite{CB}, cancer survival \cite{Chena}, and econophysics \cite{MK}. The
application of the superstatistics to RMT \cite{Magd1} assumes the spectrum of
a mixed system as made up of many smaller cells that are temporarily in a
chaotic phase. Each cell is large enough to obey the statistical requirements
of RMT but has a different distribution parameter $\eta$ associated with it,
according to a probability density $f(\eta)$. Consequently, the
superstatistical random matrix ensemble that describes the mixed system is a
mixture of Gaussian ensembles. \ The joint probability density distribution of
the matrix elements is obtained by integrating distributions of the form%
\begin{equation}
P(H)=\frac{1}{Z(\eta)}exp[-\eta Tr(H^{\dag}H)] \label{rmt}%
\end{equation}
over all positive values with a statistical weight $f(\eta)$, which leads to
\begin{equation}
P(H)=\int_{0}^{\infty}f(\eta)\frac{exp[-\eta Tr(H^{\dag}H)]}{Z(\eta)}%
d\eta\label{sust}%
\end{equation}
where $Z(\eta)=\int exp[-\eta Tr(H^{\dag}H)]dH,$

One of the fundamental assumptions of RMT is that the matrix element joint
probability density distribution is base independent. This makes the theory
suitable for modelling quantum chaotic systems. Indeed, the eigenfunctions of
a Hamiltonian of a system with a chaotic classical limit are unknown in
principle. In other words, there is no special basis to express the
eigenstates of a chaotic system. In integrable systems, on the other hand, the
eigenbasis of the Hamiltonian is known in principle. In this basis, each
eigenfunction has just one component that obviously indicates the absence of
complexity. In the nearly ordered regime, mixing of quantum states belonging
to adjacent levels can be ignored and the energy levels are uncorrelated.

\subsection{Parameter distribution}

The distribution $f(\eta)$ is determined by the spatiotemporal dynamics of the
entire system under consideration. Beck et al.\cite{CBS} have argued that
typical experimental data are described by one of three superstatistical
universality classes, namely, $\chi^{2}$, inverse $\chi^{2}$ or log-normal
superstatistics. The first universality holds if $\eta$\ has contribution from
$\nu$ Gaussian random variables $X_{1}$, $...$ , $X_{\upsilon}$ due to various
relevant degrees of freedom in the system. Then a positive $\eta$ is obtained
by setting $\eta=\sum_{i=1}^{\upsilon}X_{i}^{2}$ is and $f(\eta/\eta_{0})$ is
a \bigskip$\chi^{2}$ distributed with degree $\nu,$%
\begin{equation}
f(\eta/\eta_{0})=\frac{1}{\Gamma(\nu/2)}(\frac{\nu}{2\eta_{0}})^{\nu/2}%
\eta^{\nu/2-1}e^{-\nu\eta/2\eta_{0}}%
\end{equation}
Here, $\eta_{0}$ is the average value of $\eta,$ with $\eta_{0}=\int
_{0}^{\infty}\eta f(\eta/\eta_{0})d\eta$

In the second universality class, the same considerations as above can be
applied if \ $\eta^{-1},$ rather than $\eta$, is the sum of several squared
Gaussian random variables. The resulting $f(\eta/\eta_{0})$ is the inverse-
$\chi^{2}$ distribution given by%
\begin{equation}
f(\eta/\eta_{0})=\frac{\eta_{0}}{\Gamma(\nu/2)}(\frac{\nu\eta_{0}}{2})^{\nu
/2}\eta^{-\nu/2-2}e^{-\nu\eta_{0}/2\eta} \label{InvChi}%
\end{equation}
In the third universality class, instead of being a sum of many contributions,
the random variable $\eta$\ may be generated by multiplicative random
processes, i.e. $\eta=%
{\displaystyle\prod\limits_{i=1}^{\nu}}
X_{i}$. Then $\ln\eta=\sum_{i=1}^{\nu}\ln X_{i}$ is a sum of Gaussian random
variables; hence it is Gaussian as well. Thus is log-normally distributed,
i.e.,%
\begin{equation}
f(\eta/\eta_{0})=\frac{1}{\sqrt{2\pi}\nu\eta}e^{-[\ln(\tfrac{\eta}{\mu}%
)]^{2}/2\nu^{2}},
\end{equation}
which has an average $\mu\sqrt{\psi}$ and variance $\sigma^{2}=$ $\mu^{2}%
\psi(\psi-1)$, where $\psi=\exp(\nu^{2}$\bigskip$)$.

\subsection{NNS distribution}

It follows from Eq.(\ref{sust}) that the statistical measures of the
eigenvalues of the superstatistical ensemble are obtained as an average of the
corresponding $\eta$-dependent ones of standard RMT weighted with the
parameter distribution $f(\eta/\eta_{0})$. In particular, the superstatistical
NNS distribution is given by \cite{Magd1} as%
\begin{equation}
p(s)=\int_{0}^{\infty}f(\eta/\eta_{0})p_{w}(\eta,s)d\eta,
\end{equation}
where $p_{w}(\eta,s)$ is the Wigner surmise for the Gaussian orthogonal
ensemble with the mean spacing depending on the parameter,%
\begin{equation}
p_{w}(\eta,s)=\eta s\exp(-\frac{1}{2}\eta s^{2}).
\end{equation}
For a $\chi^{2}$ distribution of the superstatistical parameter $\eta$ the
resulting NNS distribution is given by%
\begin{equation}
p_{\chi^{2}}(\nu,s)=\frac{\eta_{0}s}{(1+\eta_{0}s^{2}/\nu)^{1+\nu/2}}.
\end{equation}
The parameter $\eta_{0}$ is fixed by requiring that the mean level spacing
$\langle s\rangle$ equals unity, yielding
\begin{equation}
\eta_{0}=\frac{\pi\nu}{4}[\Gamma(\frac{\nu-1}{2})/\Gamma(\frac{\nu}{2})]^{2}.
\end{equation}
For an inverse $\chi^{2}$ distribution, one obtains%
\begin{equation}
p_{inv\chi^{2}}(\nu,s)=\frac{2\eta_{0}s}{\Gamma(\frac{\nu}{2})}(\sqrt{\eta
_{0}\nu s/2})^{\nu/2}K_{\nu/2}(\sqrt{\eta_{0}\nu s}), \label{PS2}%
\end{equation}
where $K_{m}(x)$ is a modified Bessel function \cite{Gradshteyn} and $\eta
_{0}$ again is determined by the requirement that the mean level spacing
$\langle s\rangle$ equals unity. Finally, if the parameter has a log- normal
distribution (5), then the NNS distribution%
\begin{equation}
p_{LogNorm}(\nu;s)=\frac{s}{\sqrt{2\pi}\nu}\int_{0}^{\infty}\exp[-\frac{\eta
s^{2}}{2}-\frac{\ln^{2}(\frac{2}{\pi}\eta e^{-\nu^{2}/4})}{2\nu^{2}}]d\eta
\end{equation}
can only be evaluated numerically.\newline

A justification for the use of the above-mentioned superstatistical
generalization of RMT in the study of mixed systems, is given in \cite{DA}. It
is based on the representation of their energy spectra in the form of discrete
time series in which the level order plays the role of time. Reference
\cite{DA} considers two billiards with mushroom-shaped boundaries as
representatives of systems with mixed regular--chaotic dynamics and three with
the shape of Lima\c{c}on billiards, one of them of chaotic and two of mixed
dynamics. The time series analysis in Ref. \cite{DA} allows to derive a
parameter distribution $f(\eta)$. The obtained distribution agrees better with
the inverse $\chi^{2}$ distribution given by Eq. (\ref{InvChi}). The inverse
$\chi^{2}$ distribution of $\eta$ follows when the quantity $\eta$\ is the sum
of $\nu$ inverse-squared Gaussian random variables. In the application to RMT,
the parameter $\eta$ is proportional to the sum of the inverse variances of
the matrix elements of $\mathbf{H}$ \cite{DA}. Hence, $\nu$ refers to the
number of (largely) contributing matrix elements. In the limit of
$\nu\rightarrow\infty$ where all the matrix-elements contribute to the
distribution, Eq. (\ref{InvChi}) yields a delta function for $p_{inv\chi^{2}%
}(\nu,s)$ turning superstatistical matrix-element distribution (\ref{sust})
into that of the conventional RMT, Eq. (\ref{rmt}). If we take this assumption
literarily, we must restrict $\nu$ to take positive integer values. As the
transition from integrability to chaos is known to proceed continuously, we
have to relax this condition and allow $\nu$ to take any real value grater
than 1. Using the asymptotic expression of the modified Bessel function
\cite{Gradshteyn}, we easily find the Wigner surmise when $\nu\rightarrow
\infty$, as required. The other limit of $\nu\rightarrow1$ yields the
semi-Poisson distribution%
\begin{equation}
p_{\text{SS}}(1,s)=4se^{-2s}, \label{SP}%
\end{equation}
which is known to provide a satisfactory description for the spectra of
pseudointegrable systems such as planar polygonal billiards, when all their
angles are rational with $\pi$ \cite{bog}. The motion of the corresponding
classical systems in phase space is not restricted to a torus like for
integrable systems, but to a surface with a more complicated topology
\cite{richens}. We therefore conclude that the assumption that the inverse
square of the variance of matrix elements as an inverse $\chi^{2}$ variable
allows superstatistical RMT to model the transition out of chaos
(corresponding to $\nu\gg1)$ until the system reaches the state of quasi-integrability.

It is interesting to note that, if one allows $\nu$ to take lower values, one
finds that the distribution $\left(  \ref{PS2}\right)  $ tends to the Poisson
distribution as $\nu\rightarrow-1$;%
\begin{equation}
p_{\text{SS}}(-1,s)=e^{-s}.
\end{equation}
We there conclude that formula (\ref{PS2}) can provide a successful model for
describing the stochastic transition all the way from integrability to chaos
passing by the stage of quasi-integrability. This has been clearly
demonstrated in Ref. \cite{DA}. We have no physical explanation for this
success. We regard $p_{\text{SS}}(\nu,s)$ in the range of $-1\leq\nu\leq1$ as
a clever parametrization of NNS distribution of nearly integrable systems
undergoing a transition from a Poisson to semi-Poisson statistics.

\section{Data analysis}

Our purpose here is to show that the superstatistical RMT is suitable for the
analysis of chaotic systems undergoing a process leading to the breaking \ of
a discrete symmetry. In our view, symmetry breaking has something in common
with the transition from integrability to chaos. The presence of a symmetry
favors the particular bases in which the eigenvectors of the symmetry are
components of the eigenstates of the Hamiltonian (e.g., the isospin
wavefunctions in the nuclear physics problem). In this representation, the
Hamiltonian matrix is block diagonal and its spectrum consists of independent
sequence of eigenvalues. As symmetry breaking interactions increase, the
eigenstates involved in the different blocks mix filling the "empty" places in
the Hamiltonian matrix. At a certain stage of the symmetry breaking
transition, the bases in which the eigenvectors of the symmetry are components
of the eigenstates of the Hamiltonian loose their special status. The joint
matrix-element distribution becomes base independent. At this point, we expect
superstatistical RMT to become suitable for describing the symmetry breaking
process with the superstatistical parameter $\nu$ measuring the number of
effective matrix elements responsible for symmetry breaking.

In the quartz crystal experiment \cite{Ellegaard q}, the authors noted that
the block with conserved flip symmetry has much in common with a scalar
pseudointegrable system \cite{bog,richens}. The initial state of the
transition can be described by an independent superposition of two independent
semi-Poissonian sequences of equal densities. Applying the method given in
Mehta's book \cite{RMT}, we obtain for NNS distribution%
\begin{equation}
p_{\text{2PI}}(s)=\frac{1}{2}e^{-2s}\left(  1+4s+2s^{2}\right)  .
\end{equation}
We have found out that the least square difference between $p_{\text{2PI}}(s)$
and the superstatistical distribution (\ref{PS2}) in the spacing interval
$0<s<3.0$,%
\begin{equation}
\int_{0}^{3.0}\left\vert p_{\text{2PI}}(s)-p_{inv\chi^{2}}(\nu,s)\right\vert
^{2}%
\end{equation}
has a minimum at $\nu=-0.210$. In the following, we show that NNS
distributions of resonances in the quartz crystal experiment \cite{Ellegaard
q}, the authors noted that the initial state of the transition can be
described by the distribution (\ref{PS2}) allover the symmetry breaking
transition by allowing $\nu$ to vary in the range of $-0.2\leq\nu\leq\infty$.

In figure 1, we compare the experimental results of the acoustic resonances
measured by Ellegaard et al. \cite{Ellegaard q} with and the superstatistical
NNS distributions corresponding to the $\chi^{2}$, the inverse-$\chi^{2}$, an
the log-normal distributions of the superstatistical parameter. We also show
in Fig. 1 the results of a previous analysis of the same data ( figure 5 in
Ref. \cite{Abd El-Hady}) with a random matrix model in which assumes that the
spectra is composed of three independent components, two pseudo-integrable
sequences for the conserved symmetry and one GOE sequence for the broken
symmetry. The best-fit values of the parameters are given in Table 1. We
quantify the quality of the fits by their absolute average deviation,
$\Delta=\frac{1}{N_{L}}\sum\left\vert P_{Exp}-P_{Cal.}\right\vert ,$ where
$P_{Exp}$, $P_{Cal},$and $N_{L}$ are the experimental, the calculated NNS
distributions and the number of measures values. The results suggests the
validity of the superstatistical distribution, even in the initial stages of
the breaking of the symmetry. The table clearly shown that the NNS
distribution obtained from the inverse-$\chi^{2}$ distribution (\ref{InvChi})
agrees with experiment data better than the other distributions even in the
initial stages of the symmetry breaking process.

\begin{center}%
\begin{tabular}
[c]{|l|l|l|l|l|l|l|l|}\hline
$r\ $ & \multicolumn{1}{|l}{El-Hady et al} & \multicolumn{2}{|l}{$\chi^{2}$
distribution} & \multicolumn{2}{|l}{inverse $\chi^{2}$ distribution} &
\multicolumn{2}{|l|}{Log normal distribution}\\\hline
(mm) & $\Delta$ & $\nu_{c}$ & $\Delta$ & $\nu_{i}$ & $\Delta$ & $\nu_{_{L}}$ &
$\Delta$\\\hline
0 & 0.034 & 3.387 & 0.063 & -0.112 & 0.035 & 1.100 & 0.053\\\hline
0.5 & 0.037 & 2.887 & 0.061 & -0.025 & 0.032 & 1.176 & 0.047\\\hline
0.8 & 0.045 & 3.121 & 0.046 & 0.405 & 0.026 & 1.064 & 0.033\\\hline
1.1 & 0.048 & 4.607 & 0.029 & 1.832 & 0.021 & 0.779 & 0.025\\\hline
1.4 & 0.056 & 6.67 & 0.023 & 4.385 & 0.024 & 0.600 & 0.023\\\hline
1.7 & 0.036 & 10.528 & 0.020 & 7.528 & 0.019 & 0.469 & 0.019\\\hline
10 & 0.029 & 12.183 & 0.026 & 9.376 & 0.026 & 0.432 & 0.026\\\hline
\end{tabular}

\end{center}

Table 1: The best fit parameters and absolute average deviations $\Delta$ for
different radii $r$\ of the removed corner of the quartz block in the
comparison of NNS distributions of acoustic resonances and the
superstatistical RMT with a $\chi^{2}$, inverse-$\chi^{2}$ and log normal
parameter distributions. The results of analysis of the three-level-sequence
model by El-Hady et al. \cite{Abd El-Hady} are also shown

\section{Summary and conclusion}

We have described the symmetry breaking of \ the acoustic resonance in a
quartz blocks, using a superstatistical model that has been successfully
applied to describe systems with mixed regular-chaotic dynamics within the
framework of RMT. Superstatistical RMT is a base independent approach and is
suitable to model symmetry breaking only when the symmetry representations
become mixed enough. Superstatistics arises by allowing the mean density of
states to fluctuate according to given distribution. We examined three
possible parameter distributions, namely the $\chi^{2}$, the inverse-$\chi
^{2}$ and the log normal distributions. Our analysis shows that the
inverse-$\chi^{2}$ distribution agrees with experimental spectra of acoustic
resonances better than the other two, in the same way as in a previous
analysis of stochastic transition of mixed microwave billiards. The
superstatistical parameter $\nu$ that characterizes the fluctuation of the
mean level density will hear measure the degree of symmetry breaking. We also
show that NNS distributions with an inverse-$\chi^{2}$ superstatistics provide
a reasonable description of experiment data not only when the system
approaches the state of chaos, but also in the initial stage of the symmetry
breaking transition when base invariance is not expected.

\bigskip

\bigskip\textbf{Figure Caption}

Figure 1. NNS distributions for different radii $r$ of the octant removed from
the quartz blocks. The experimental data reported in \cite{Ellegaard q} are
shown by histgrams. The solid lines are the superstatistical results calculted
with an inverse-$\chi^{2}$ parameter distribution. Results for the $\chi^{2}$
and log normal parameter distributions are shown by dotted and dashed dotted
dotted crves, respectively. The dashed curves are calculated with the
three-level-sequence model by El-Hady et al. \cite{Abd El-Hady}.

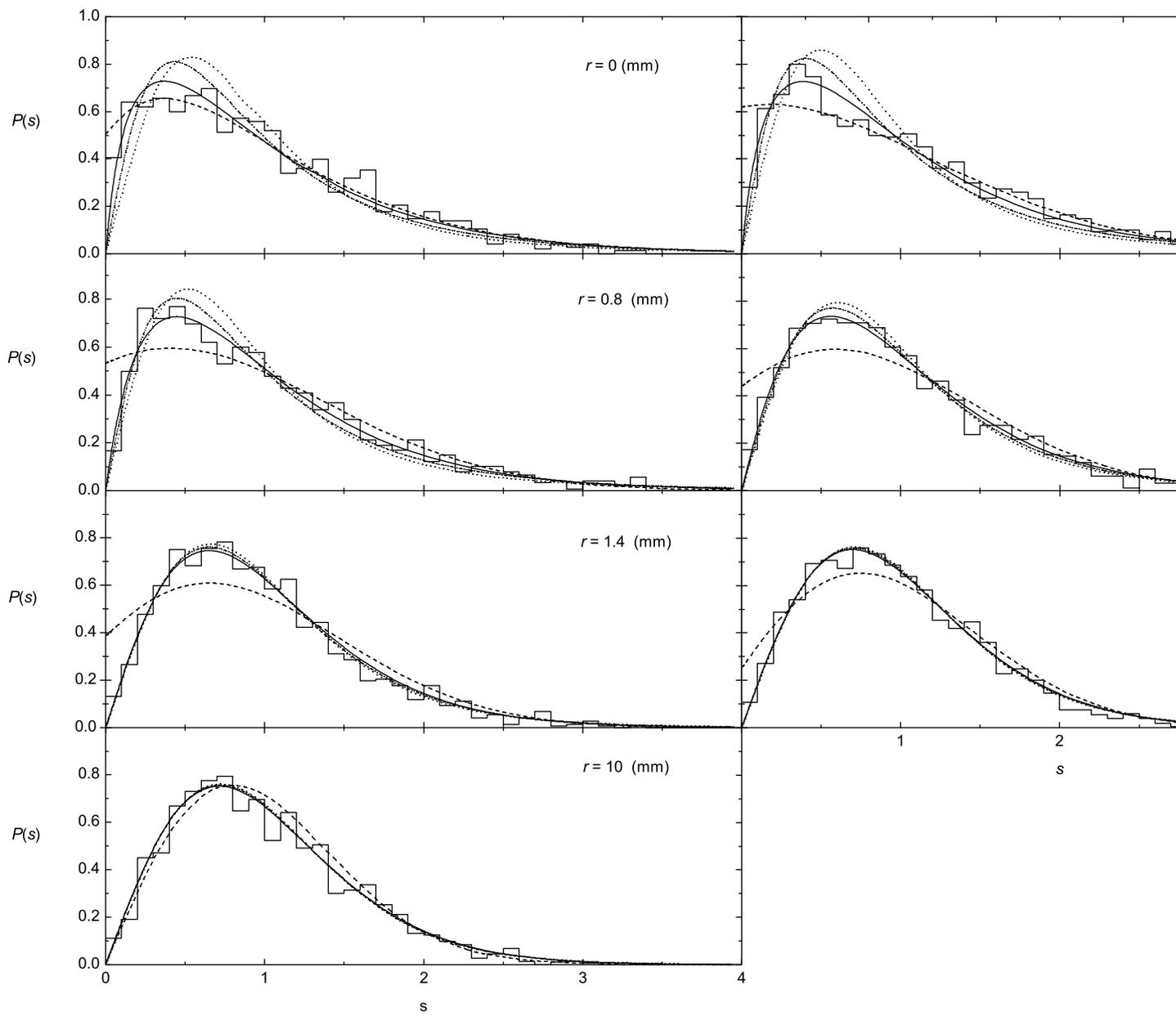
\end{document}